\documentclass[fleqn,usenatbib]{mnras}

\usepackage{newtxtext,newtxmath}
\usepackage[T1]{fontenc}
\usepackage{ae,aecompl}

\usepackage{graphicx}	% Including figure files
\newcommand{\new}[1]{#1}
\newcommand{\newb}[1]{#1}
\newcommand{\newc}[1]{#1}
\newcommand{\kms}{km\,s$^{-1}$}
\newcommand{\Teff}{$T_{\rm eff}$}
\newcommand{\logg}{$\log\,g$}
\newcommand{\vsini}{$v\sin{i}$}
\newcommand{\vmic}{$v_{\rm mic}$}
\title[KELT-9 parameters and planet evaporation]{KELT-9 and its ultra-hot Jupiter: stellar parameters, composition, and planetary pollution}
\author[Kama et al.]{Mihkel Kama$^{1,2}$\thanks{Correspondence: \texttt{m.kama@ucl.ac.uk}}, Colin P. Folsom$^{2}$, Adam S. Jermyn$^{3}$, Johanna K. Teske$^{4}$\\
$^{1}$Department of Physics and Astronomy, University College London, Gower Street, London, WC1E 6BT, UK\\
$^{2}$Tartu Observatory, Observatooriumi 1, T\~{o}ravere 61602, Tartu, Estonia\\
$^{3}$Kavli Institute for Theoretical Physics, University of California at Santa Barbara, Santa Barbara, CA 93106, USA\\
$^{4}$Earth and Planets Laboratory, Carnegie Institution for Science, 5241 Broad Branch Road, NW, Washington, DC 20015, USA}
\date{Accepted XXX. Received YYY; in original form ZZZ}

\pubyear{2022}

\begin{document}
\label{firstpage}
\pagerange{\pageref{firstpage}--\pageref{lastpage}}
\maketitle

\begin{abstract}
KELT-9b is an ultra-hot Jupiter observed to be undergoing extreme mass loss. Its A0-type host star has a radiative envelope, which makes its surface layers prone to retaining recently accreted material. To search for potential signs of planetary material polluting the stellar surface, we carry out the most comprehensive chemical characterisation of KELT-9 to-date. New element detections include Na and Y, which had previously been detected in the ultra-hot Jupiter but not studied in the star; these detections complete the set of \newc{nine }elements measured in both star and planet. In comparing KELT-9 with similar open cluster stars we find no strong anomalies. This finding is consistent with calculations of photospheric pollution accounting for stellar mixing and using observationally estimated KELT-9b mass loss rates. We also rule out recent, short-lived intensive mass transfer such as the stellar ingestion of an Earth-mass exomoon.
\end{abstract}


%%%%%%%%%%%%%%%%%%%%%%%%%%%%%%%%%%%%%%%%%%%%%%%%%%

%%%%%%%%%%%%%%%%% BODY OF PAPER %%%%%%%%%%%%%%%%%%

\section{Introduction}

Accretion of planetary material can potentially affect the photospheric composition of the host star. This can occur due to engulfment \citep{Israelianetal2001, Nagaretal2020} or transfer of mass by a planetary wind \citep{Jura2015} and has the potential to offer diagnostic power complementary to methods like transit spectroscopy. We analyse a high \newc{signal-to-noise ratio (SNR) }spectrum of KELT-9 (HD~195689), which hosts the hottest and most rapidly evaporating planet found to-date\new{, with the aim of looking }for composition anomalies \new{caused by accretion of planetary material}.

\begin{figure*}
\centering
\includegraphics[clip=,width=1.0\linewidth]{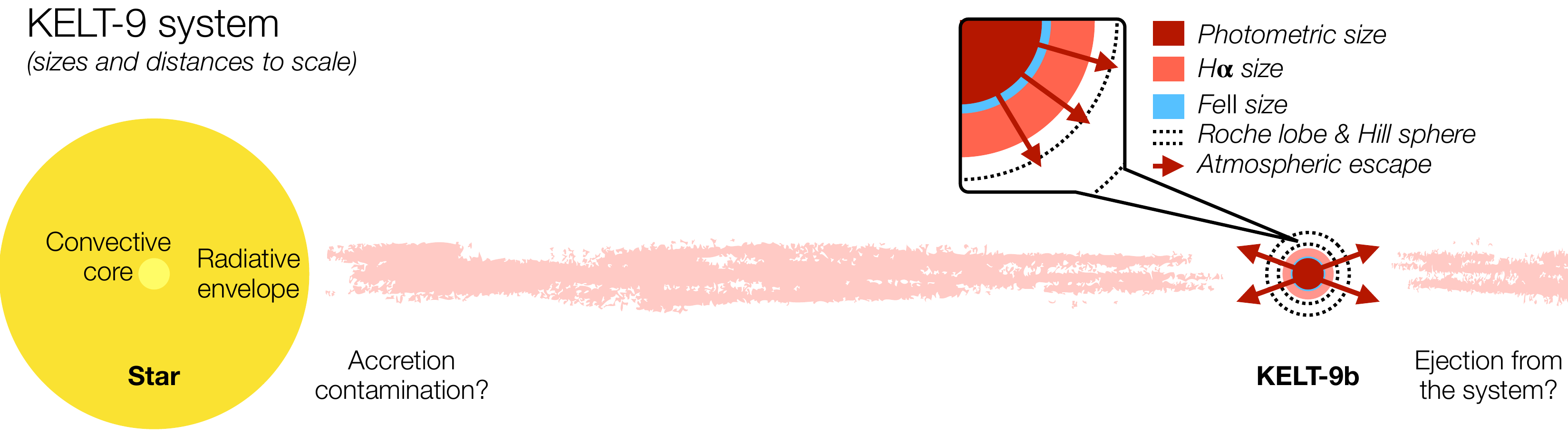}
\caption{KELT-9b and its host star to scale. The composition of the evaporated planetary material and the mass loss rate, and the capture efficiency of the planetary wind material by the central star, may be constrained by stellar spectroscopy.}
\label{fig:cartoon}
\end{figure*}

KELT-9b is an intensely irradiated ultra-hot Jupiter around an A0-type star \citep{Gaudietal2017}. Orbiting at $a_{\rm maj}=0.035\,$au from its star with a period of $P_{\rm orb}=1.48\,$days, KELT-9b has a mass of $2.88\pm0.84\,$M$_{\rm Jup}$, radius of $1.891^{+0.061}_{-0.053}\,$R$_{\rm Jup}$, and an equilibrium temperature of $4050\pm180\,$K with $T_{\rm day}\approx4600\,$K measured on the dayside \citep{Gaudietal2017}. Atmospheric mass loss \new{ measurements range from $1-3\times10^{12}\,$g\,s$^{-1}$ \citep{YanHenning2018, Cauleyetal2019, Wyttenbachetal2020}}. \newc{A recent result using NLTE calculations finds a pure oxygen loss rate of $10^{8}$ to $10^{9}\,$g\,s$^{-1}$ \citep{Fossatietal2020, Fossatietal2021, Borsaetal2022}. }The observational estimates \newc{for the bulk mass loss mostly }fall within the theoretical UV-flux limited rate of $10^{10}$ to $10^{13}\,$g\,s$^{-1}$ \citep[$10^{-16}$ to $10^{-13}\,$M$_{\odot}$\,yr$^{-1}$; see][]{Gaudietal2017, Fossatietal2018}\newc{, where the upper end may be based on a stellar extreme-UV luminosity that is too high for A-type stars}. \new{A to-scale diagram of KELT-9 and its evaporating planet is shown in Figure\,\ref{fig:cartoon}.}

For stars of mass ${\gtrsim}1.4\,$M$_{\odot}$ (spectral type F5), which have slowly-mixing radiative envelopes, accretion can dominate the composition of the surface more easily than for stars with \newb{large }convective envelopes \citep{CharbonneauMichaud1991, TurcotteCharbonneau1993, Turcotte2002, JermynKama2018}. This \new{characteristic }makes mid-F and A-type stars interesting candidates to seek signs of recent or ongoing ingestion of planetary material.

\section{Observations}
We obtained a Gemini/GRACES spectrum of HD\,195689 (KELT-9) on September 13th, 2017, through proposal GN-2017B-DD-1. GRACES, or the Gemini Remote Access to CFHT ESPaDOnS Spectrograph \citep{Tollestrupetal2012, Cheneetal2014}, combines the collecting power of the 8.1m Gemini North telescope with the resolving power and efficiency of the ESPaDOnS spectrograph on the CFHT through a 270m fiber feed. We used the 1-fiber mode, which provides a maximum spectral resolution of $R\approx67,500$. \newb{This fixed instrument configuration provides nearly continuous spectra from ${\approx}4000$ to $10400\,$\AA\ in 35 echelle orders. }During our observations the average airmass and seeing were 1.1 and $0.55''$, respectively, and we used 16x60s consecutive exposures to achieve a \newc{SNR${\approx}590$ }per $1.8\,$km\,s$^{-1}$ spectral pixel around $6000\,$\AA\ (and \newc{SNR${\approx}500$ }at $5000\,$\AA). The data were reduced with \new{both the }DRAGRACES \citep{Cheneetal2021} and OPERA \new{extraction pipelines }and cross-checked. At wavelengths $<6500$\,\AA\ DRAGRACES resulted in a better SNR, \new{while }at longer wavelengths the \newc{SNR }of the reductions were consistent but OPERA had a smoother continuum so it was used in analysing the longest wavelength window.

\section{Determining the stellar properties and composition}

In order to discuss any potential anomalies in the composition of KELT-9, such as signatures of contamination by mass lost from its ultra-hot Jupiter, we first determine the stellar parameters and photospheric chemical element abundances.

The analysis closely followed the approach of \citet{Folsometal2012}.  The approach was to directly fit the observations with model spectra, simultaneously deriving abundances and stellar parameters through $\chi^2$ minimization.  We fit six large spectral windows independently and use the average and standard deviation of results from these window to, approximately, account for systematic uncertainties.  To calculate synthetic spectra we used the {\sc Zeeman} spectrum synthesis code \citep{Landstreet1988, Wade2001}, with a Levenberg-Marquardt fitting routine and optimizations for a negligibly magnetic star \citep{Folsometal2012}.  {\sc Zeeman} performs polarized radiative transfer in LTE including the Zeeman effect and produces model spectra integrated across the visible disk of a star.  A grid of model atmospheres from ATLAS9 \citep{Kurucz1993-ATLAS9,Castelli2003-ATLAS9} were used as input, and a bi-linear interpolation of the log of the model quantities was used to produce an atmosphere for a specific \Teff\ and \logg.  Atomic line data was taken from the Vienna Atomic Line Database \citep[VALD][]{Piskunov1995-VALD,Ryabchikova1997-VALD,Kupka1999-VALD,Kupka2000-VALD,Ryabchikova2015-VALD}, using an `extract stellar' request for the parameters of KELT-9.  

The free parameters in the fit were \Teff, \logg, \vsini, microturbulence (\vmic), radial velocity, and chemical abundances ($\log_{10}{(X/H)}$).  We assumed an instrumental resolution of 65000 for GRACES.  The magnetic field and the macroturbulence, which is likely $\ll$\,\vsini, were assumed to be negligible.

We normalized the observed spectrum by fitting low degree polynomials through carefully selected continuum points in individual spectral orders.  Each spectral order was normalized independently then merged, rejecting the lower SNR order edges, to produce a continuous well normalized observation.  

We selected six large spectral window that were fit independently: 4147--4292.5 \AA, 4406--4785 \AA\ (excluding the Mg {\sc ii} 4481 \AA\ line), 4999--5500 \AA, 5500--6000 \AA\ (excluding the Na D doublet and stronger telluric H$_2$O lines between $\sim$5885 and $\sim$5980 \AA), 6000--6460 \AA\ (excluding the strong telluric O$_2$ band between 6275 and 6330 \AA), and portions of 6660--7590 \AA\ with little telluric contamination (specifically 6660--6860 + 7085--7130 + 7400--7590 \AA).  These windows were chosen to span most of the available spectrum where the SNR is good.  Balmer lines were avoided since their precise normalization in echelle spectra is unreliable.  The particularly strong Mg {\sc ii} 4481 \AA\ line was excluded since it produced results discrepant with all other Mg lines in the spectrum.  Large windows were used to provide many lines with a range of ionization, excitation potential, and strength, in order to simultaneously constrain stellar parameters and abundances. The exact window edges were adjusted to fully include lines but exclude potential sources of systematic error.  

\begin{figure*}
\centering
\includegraphics[trim={0.0cm 0.0cm 0.0cm 0.0cm}, clip=,width=1.0\textwidth]{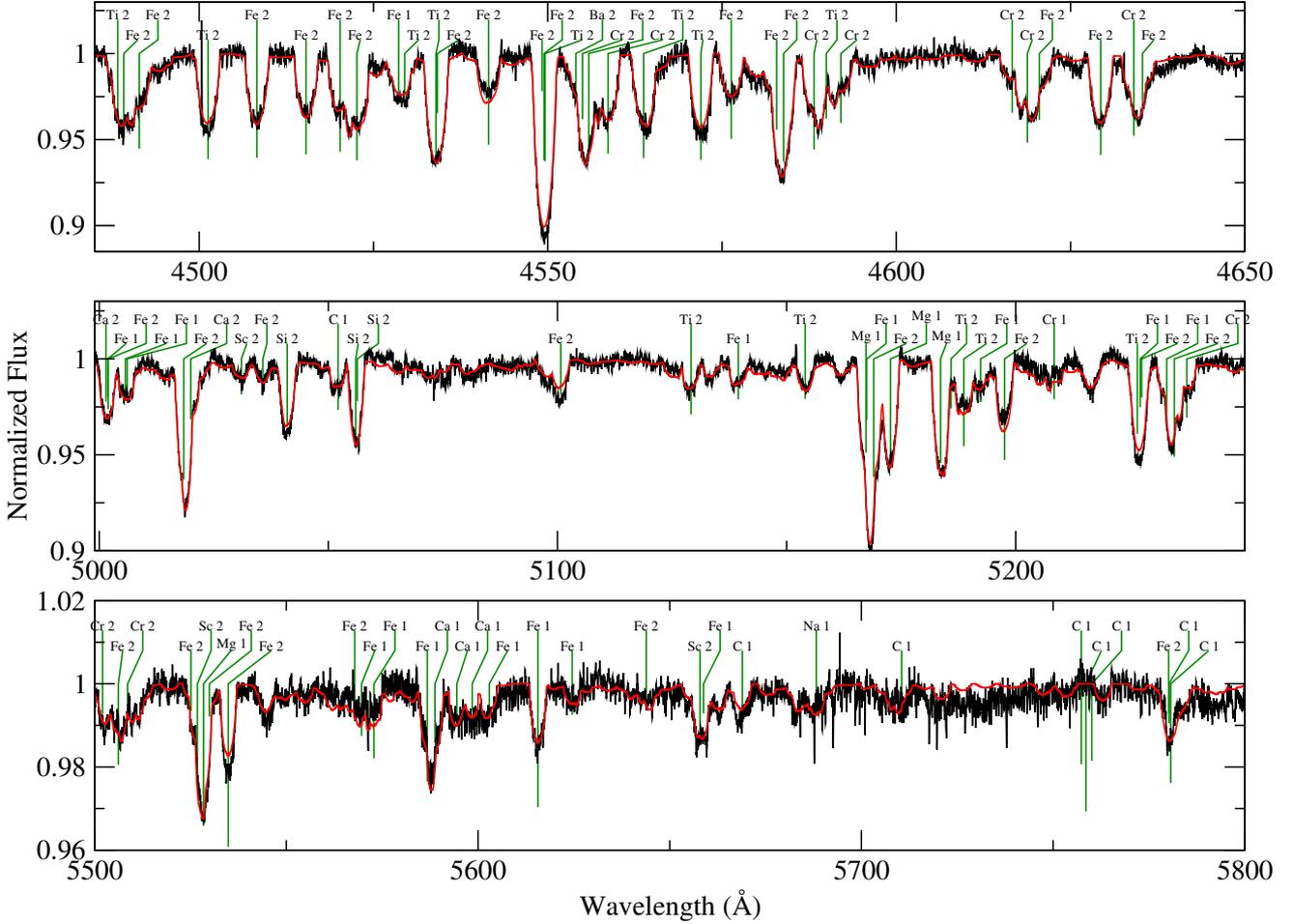}
\caption{Selected regions of the spectrum of KELT-9, showing the GRACES data (black) and our best-fit model (red). Significant lines are labelled.}
\label{fig:specchunks}
\end{figure*}

\newb{Sections of the observed and best-fit spectra are shown in Figure\,\ref{fig:specchunks}. }The specific elemental abundances included as free parameters in the fit for each window depended on the presence of useful lines from that element.  Lines were first identified by comparing the VALD line list with features in the observation, with the help of the estimated line depths provided by the `extract stellar' feature of VALD.  Elements with lines that were visible in the observation beyond the noise level were included, unless they were the substantially weaker component in the blend.  An element whose only line was part of an similar strength blend was included in the initial fit, with a note to scrutinize the result.  An initial fit was then performed with the initial selection of element abundances, as well as \Teff, \logg, \vsini, and \vmic\ as free parameters.  The fit was checked by eye to ensure good quality and validate uncertain elements.  The resulting abundances were checked for anomalous values, or anomalously large formal uncertainties (as measured by the diagonal of the covariance matrix).  In cases where an element was found to not be reliably constrained, the element was rejected as a free parameter for the window, a solar abundance \citep{Asplundetal2009} was assumed for the purpose of spectrum synthesis, and the fit to that window was repeated with the abundance fixed.

\begin{figure*}
\centering
\includegraphics[trim={0.0cm 0.0cm 0.0cm 0.0cm}, clip=,width=1.0\textwidth]{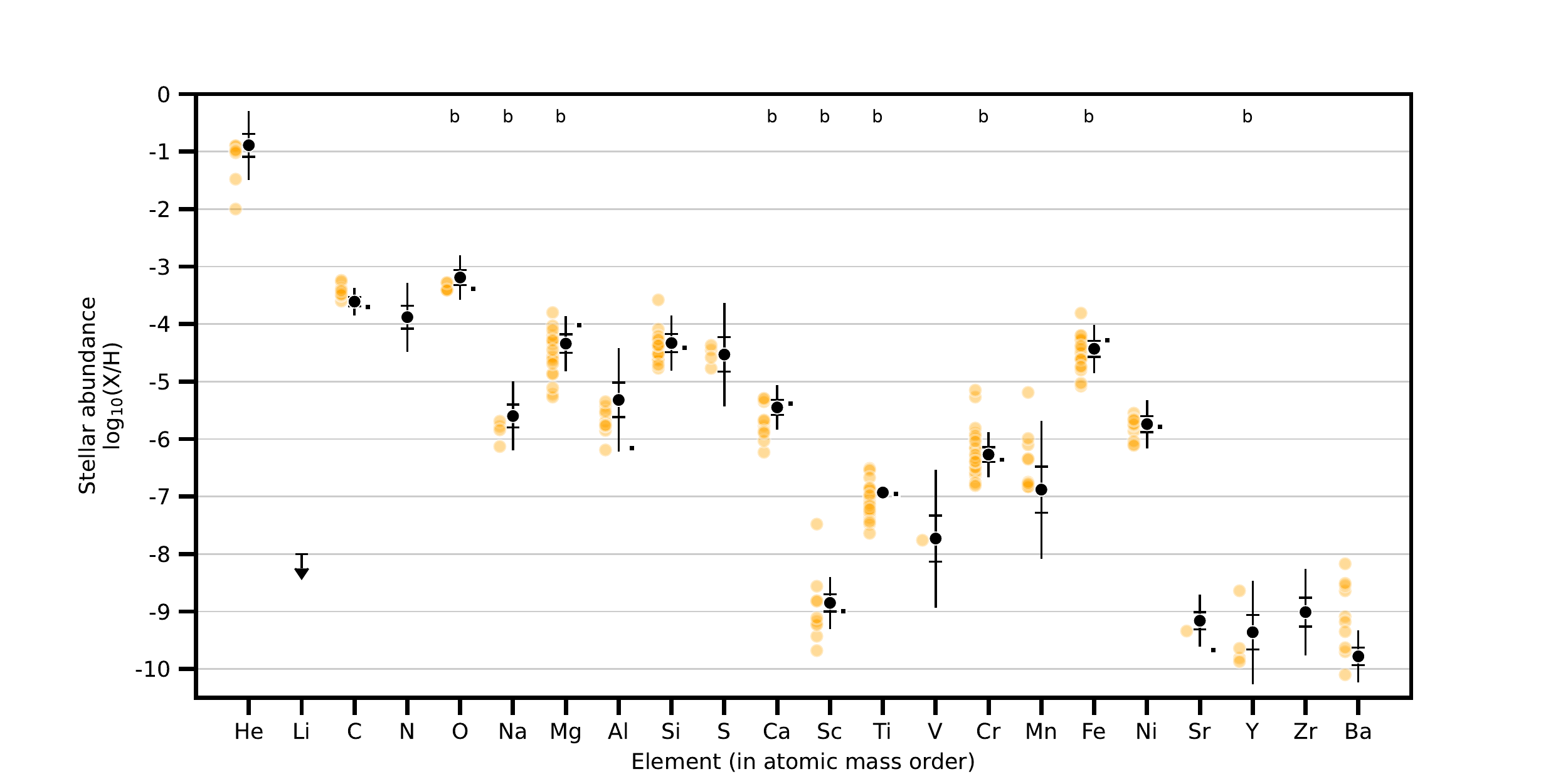}
\caption{Composition of KELT-9 (black circles with $1\sigma$ and $3\sigma$ errorbars, and $3\sigma$ Li upper limit, this work) compared to similar \Teff\ stars in the open cluster NGC\,5460 \citep[\new{orange circles,}][]{Fossatietal2011}. \new{Small black }dots show previous measurements by \citet{Saffeetal2021}. Elements detected in the atmosphere of the planet KELT-9b are marked with a `b' (see Table\,\ref{tab:param-abuns} for references).}
\label{fig:abuns}
\end{figure*}

After good fits to all six windows were achieved, a final consistency check was performed on abundances from different windows.  In cases where an abundance for one window was particularly discrepant, the reliability of that abundance was reassessed.  In cases where the discrepant abundance was based on a particularly weak line near the noise level, or a weaker component of a blended line, it was excluded from the final average.  

We found that \logg\ was poorly constrained in the  6660--7590\,\AA\ window, providing results that were clearly inconsistent with the other five windows.  This window has relatively few lines, most of which are weak, and relatively low SNR, thus the simultaneous determination of \Teff\ and \logg\ is particularly venerable to small errors in atomic data, line blending, or continuum normalization.  For the final results we fixed \logg\ for this window to the average value from the other five windows and fit for \Teff\ and the other parameters.

The final abundances and stellar parameters are presented in Table\,\ref{tab:param-abuns} and Figure\,\ref{fig:abuns}.  These are averages of the results from individual windows. The uncertainties are the standard deviation of results from the windows, except for abundances from three or fewer windows, to approximately account for systematic errors.  If errors are normally distributed, the standard error (standard deviation divided by the square root of the number of measurements) provides a better estimate on the uncertainty of the mean.  Thus the uncertainties may be over-estimated, although the standard error can be calculated easily from the quantities in Table \ref{tab:param-abuns}.  However, the sampling of atomic data from different spectral windows may not fully account for all systematic errors, and the number of windows provides small number statistics, so in some cases the uncertainties may be under-estimated.  Thus we consider the standard deviation a reasonable approximation of the real uncertainty.

\begin{figure}
\centering
\includegraphics[trim={0.0cm 0.0cm 0.0cm 0.0cm}, clip=,width=1.0\columnwidth]{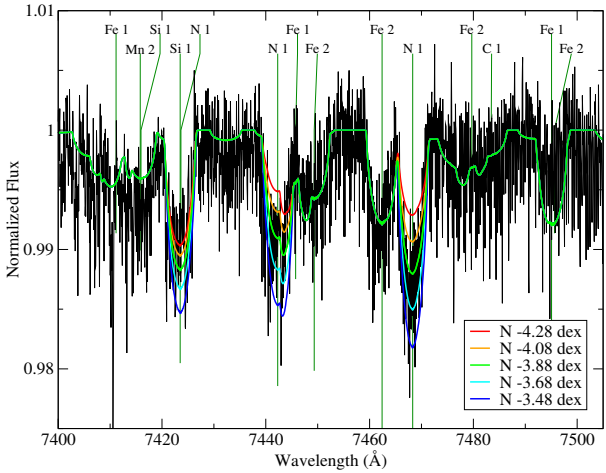}
\caption{Determination of the nitrogen abundance in KELT-9.}
\label{fig:fitN}
\end{figure}

For elements with abundances from three or fewer windows, uncertainties were estimated by visually comparing models with modified abundances to the observation. \newb{An example (N/H) is shown in Figure\,\ref{fig:fitN}. }The estimates were chosen to span the range of results from different lines, and included possible errors due to continuum normalization or blended lines.  Some of these uncertainties may be over-estimates, however the abundances may also be regarded as less reliable or more vulnerable to systematic errors.

\newc{Our stellar parameters match those in previous analyses \citep{Borsaetal2019, Saffeetal2021}. }In comparison to a previous study of the composition of KELT-9, carried out with HARPS-N data \citep{Saffeetal2021}, our wavelength coverage extends further into the red by $1600\,$\AA\ and we obtain abundances for more elements. New elements include He, N, V, Mn, Zr, and Ba, as well as Na and Y which were previously found in the extended atmosphere around KELT-9b \citep{Hoeijmakersetal2019}. We also report a conservative upper limit on Li.

\begin{figure}
\centering
\includegraphics[trim={0.0cm 0.0cm 0.0cm 0.0cm}, clip=,width=1.0\columnwidth]{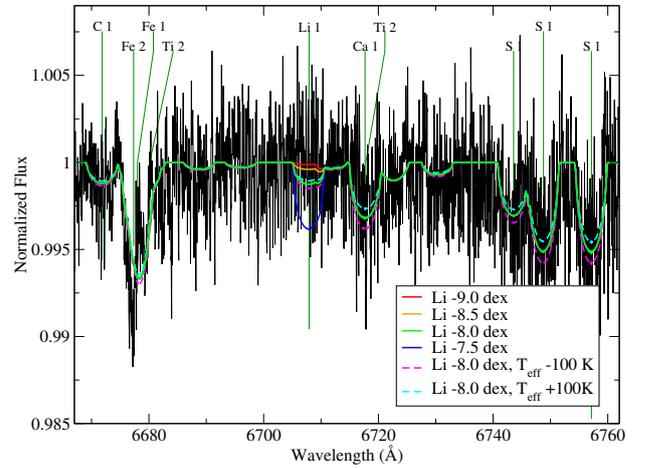}
\caption{Determination of the Li/H upper limit in KELT-9.}
\label{fig:fitLi}
\end{figure}

The Li analysis relies on the Li {\sc i} line at 6708\,\AA, and since it is not detected we can only place an upper limit on the abundance \newb{(Figure\,\ref{fig:fitLi})}. We obtain a formal limit by comparing the $\chi^2$ for a model with no Li to $\chi^2$ for models with increased Li abundances, looking for a change in $\chi^2$ that is statistically worse than the null model.  From this process we find a Li abundance $<-8.5$ dex for a $3\sigma$ confidence level.  However, this neglects any potential systematic errors.  If we consider possible continuum normalization errors comparable to the noise level, which is plausible due to the high \vsini, then by visually comparing models to the observation we only confidently find a Li abundance $<-8.0$ dex.  A change in \Teff\ by $\pm100$ K changes the abundance limit by roughly $\pm0.1$ dex.  We adopt a cautious $<-8.0$ dex limit on Li/H, but the real limit is likely between $-8.5$ and $-8.0$.

\section{Analysis}

As depicted in Figure~\ref{fig:cartoon}, our aim is to constrain any accretion stream coming from the planet and reaching the stellar surface by examining the composition of the star's photosphere.
To do this, we follow~\citet{JermynKama2018} and introduce the parameter $f_{\rm ph}$, which is the mass fraction of the photosphere composed of accreted material.
That is,
\begin{align}
	\left(\frac{\rm X}{\rm H}\right)_{\rm photosphere} = f_{\rm ph} \left(\frac{\rm X}{\rm H}\right)_{\rm accreted} + \left(1-f_{\rm ph}\right) \left(\frac{\rm X}{\rm H}\right)_{\rm native},
\end{align}
where by native we mean the composition of the material in the star prior to accretion \new{from the planet}.

We compute $f_{\rm ph}$ by comparing the time-scale over which material in the photosphere mixes with material deeper in the star to that over which new material enters the photosphere~\citep{JermynKama2018}. We do this assuming that KELT-9 is a typical main-sequence star characterised by temperature $T_{\rm eff} = 10,170{\rm K}$, mass $M=2.52M_\odot$, radius $R = 2.362R_\odot$ and rotation speed $v\sin i = 114.3 {\rm km\,s^{-1}}$~\citep{Gaudietal2017}. The uncertainties on these parameters are small compared with uncertainties in the mixing processes at work so we neglect them. We assume $\sin i = 1$ for simplicity, producing an upper bound on the rotational mixing and therefore a lower bound on $f_{\rm ph}$.

\begin{figure}
\includegraphics[width=0.46\textwidth]{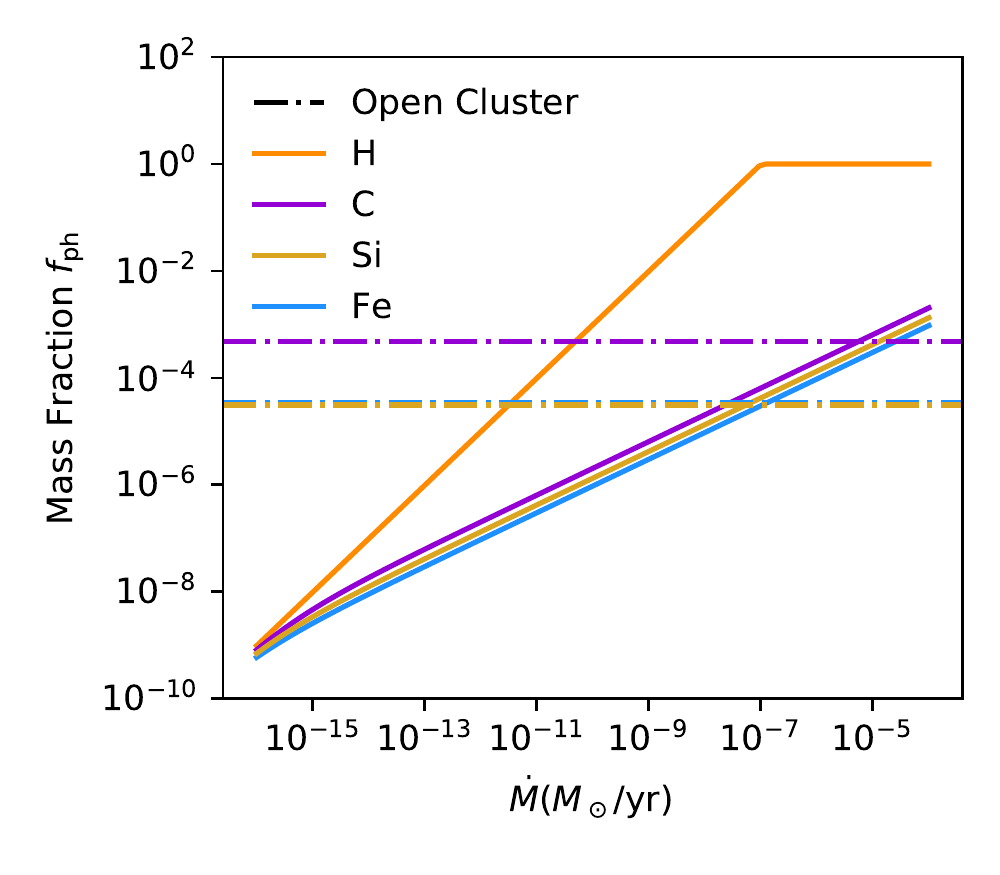}
\caption{The fractional number abundance $f_{\rm ph}$ of chemical elements in the photosphere is shown with solid lines as a function of $\dot{M}$ for pure accretion of hydrogen (H, orange), carbon (C, purple), silicon (Si, yellow), and iron (Fe, blue). The reference stellar abundances (dash-dotted lines), are inferred from a sample of open cluster stars and listed in Table\,\ref{tab:param-abuns}.}
\label{fig:fph}
\end{figure}

\begin{figure}
\centering
\includegraphics[clip=,width=1.0\linewidth]{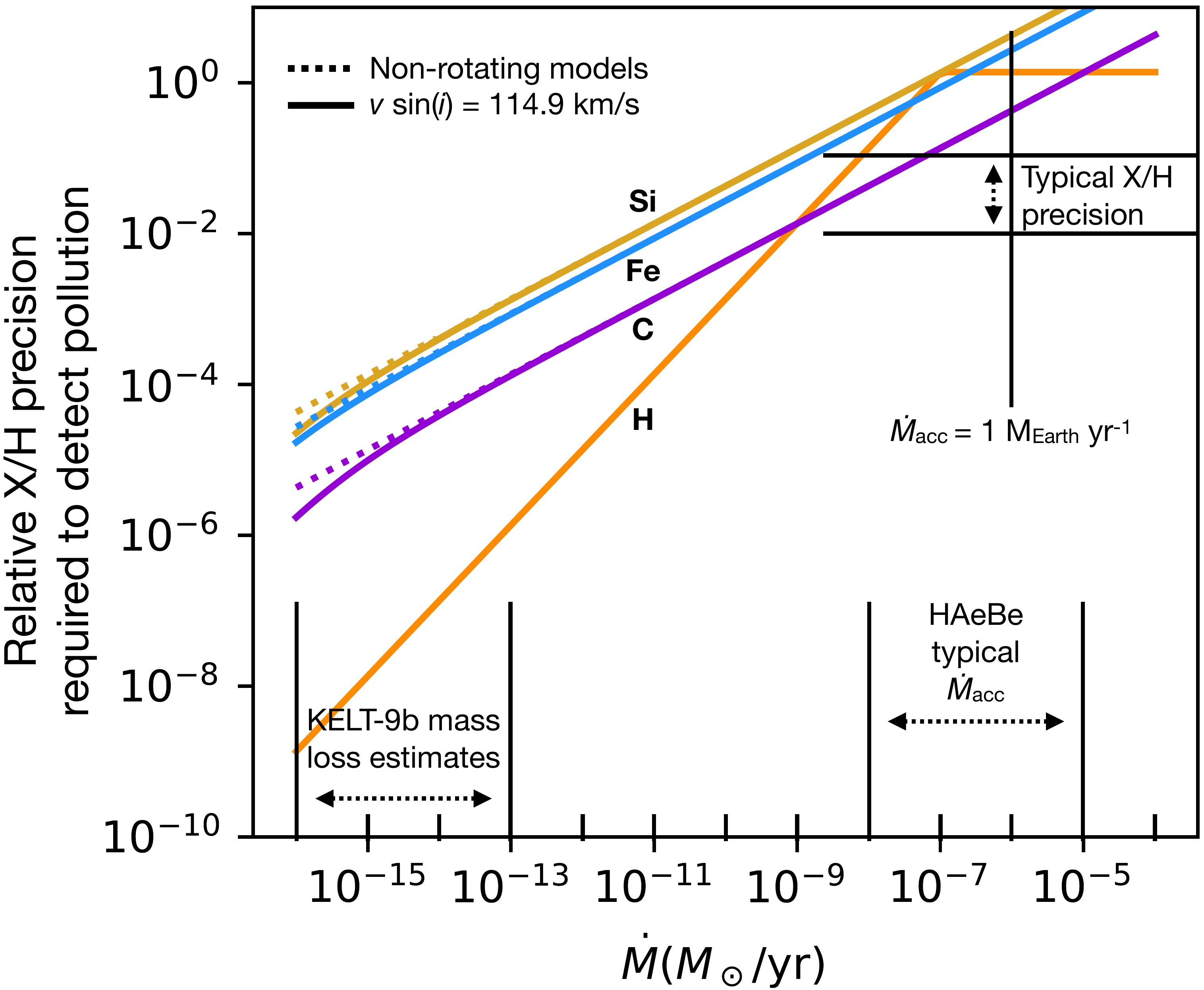}
\caption{The observational precision required to detect an accretion contamination signature on KELT-9, as a function of accretion rate and assuming different chemical elements dominating the material.}
\label{fig:mdot_sensitivity}
\end{figure}

The results are shown in Figure~\ref{fig:fph}. Because thermohaline mixing occurs at a rate \new{that }depends on the mean molecular weight of the accreted material, we sample the two extreme possibilities of pure hydrogen and pure iron, as well as \newc{the intermediate cases of pure carbon and silicon}.
For reference the abundances \newc{C ,Si, and Fe }inferred from a sample of comparable open-cluster stars are shown as well. For $f_{\rm ph}$ below these abundances the chemical signatures of accreting material should be difficult to detect unless the composition of accreting material is highly unusual.
Thus these horizontal lines should be interpreted as setting effective lower bounds on the detectable $f_{\rm ph}$. \new{The calculations }hence suggest that we should only detect accretion if the planetary mass loss is $\dot{M}_{\rm p} \ga 10^{-7}\,$M$_\odot$\,yr$^{-1}$, which \new{is impossible to achieve with planet evaporation except in } a time-limited, catastrophic event. In Figure\,\ref{fig:mdot_sensitivity}, we cast this in the form of measurement sensitivity required to detect a composition \newb{(in terms of ratios of elemental number abundances) }anomaly as a function of mass accretion rate onto the star. \newb{We define ``anomaly'' as a deviation from the mean of similar \Teff\ stars in three open clusters as listed in Table\,\ref{tab:param-abuns}.}

\new{For a high-mass planet such as KELT-9b, we expect the heavy element content to be enhanced over that of the star by perhaps a factor of five to ten \citep{Thorngrenetal2016}. \newc{Such an enhanced iron abundance has indeed been inferred for KELT-9b \citep{Pinoetal2020}. This }might \newc{be }measurable if the star was accreting pure planetary material at a high rate (see Figure\,\ref{fig:fph}). \newb{Unfortunately the mixing ratio of accreted material in the stellar photosphere is unfavourable because, for a planetary mass loss rate $\dot{M}_{\rm p}=10^{-13}\,$M$_{\odot}$\,yr$^{-1}$ and a} $100\,$\% transfer efficiency onto the star, the mass fraction of KELT-9b material in the photosphere of KELT-9 would be $\log(f_{\rm ph})=-5.98$ if accreting hydrogen-dominated material, or $-7.52$ if iron-dominated. The fact that \newb{we find }no significant peculiarity in any element or group of elements in KELT-9 is thus consistent with our calculations of $f_{\rm ph}$ which suggest the planetary evaporation rate is insufficient by four to five orders of magnitude to significantly pollute the stellar surface.}

\subsection{Constraining the recent ingestion of an exomoon}

The stellar ingestion of a large, Earth-mass exomoon \newb{detached from KELT-9b and }consisting primarily of refractory elements \newb{would provide a temporary source of very high effective accretion rate. This }constitutes a testable hypothesis for our data. \new{We caution that orbital stability considerations disfavor the retention of Earth-mass exomoons at the orbital distance of KELT-9 at the ${\sim}500\,$Myr age of the system i.e., such a moon should have been ingested at a fraction of the system's current age \citep{BarnesOBrien2002}. We therefore examine this scenario mostly as an instructive case for future studies. }The accretion of an Earth-mass moon within a year would provide $\dot{M}_{\rm p} = 3\times10^{-6}\,$M$_\odot$\,yr$^{-1}$. As Earth's iron content is $32\,$wt\% \citep[$15\,$\% atomic fraction;][]{McDonoughSun1995}, we can simplify this to assume accreting a large rocky moon is equivalent to temporarily accreting ${\sim}10^{-6}\,$M$_\odot$\,yr$^{-1}$ of pure Fe. Tracing this in Figure\,\ref{fig:fph}, we see that the photospheric iron abundance would be increased by over $100\,$\% \new{and yield a measurable effect in }such an event. 

\newb{Accretion rates $10^{-8}$ to ${\sim}10^{-6}\,$M$_\odot$\,yr$^{-1}$ overwhelm the stellar photospheric composition. This has been confirmed observationally for Herbig\,Ae/Be (``HAeBe'') stars, which are young early-type stars with rapidly accreting protoplanetary disks. Such stars exhibit a correlation of photospheric refractory element abundance with the availability of dust in the inner disk \citep{Kamaetal2015b}. }These findings are consistent with calculations such as those shown in Figure\,\ref{fig:mdot_sensitivity}. Recent studies have confirmed that anomalous HAeBe photospheric compositions are indeed limited to a polluted surface layer \citep{Murphyetal2021}. Once accretion stops, \new{the contamination signature disappears from the stellar atmosphere on approximately the same timescale as the duration of the accretion \citep{JermynKama2018}, one year in the above example where we arbitrarily chose to accrete \newc{the whole }$1\,$M$_{\oplus}$ within $1\,$yr.}

From Figure\,\ref{fig:mdot_sensitivity}, we see that Fe-rich material could be measurable on the star for ongoing accretion at rates as low as $\dot{M}_{\rm lim}=10^{-10}\,$M$_\odot$\,yr$^{-1}$. A disrupted Earth-mass moon could supply such a rate for up to $\sim10^{4}\,$yr. We can conclude that much depends on the fate of disrupted exomoon material. If the material accretes onto the star, the lifetime of the accretion disk containing the destroyed exomoon will control whether or not the signature is detectable over a significant amount of time. Our data in Figure\,\ref{fig:abuns} rule out a significant enhancement of refractory elements, and thus limit the potential stellar ingestion of any Earth-sized moons of KELT-9b to have ended $10^{4}\times\min{[1,\dot{M}_{\rm lim}/(1\,{\rm M}_{\oplus}/t_{\rm acc})]}$ years ago, \newb{where $t_{\rm acc}$ is the time needed for the exomoon material to fall onto the star.}

\begin{table}
\centering
\caption{KELT-9 (HD\,195689) stellar properties and chemical element abundances (columns 1-3, \newc{$1\sigma$ uncertainties}; this work), matching atomic species detected in the planet KELT-9b (column 4; see \emph{Notes}), \newb{and the mean abundance and min-to-max range for open cluster stars with a similar \Teff\ \citep[columns 5-6; data from][]{Fossatietal2011, Martinetal2017}.}  Abundances are provided as $\rm\log{(X/H)}$, and superscripts on the uncertainties give the number of spectral windows used. Velocities are in \kms.}
\begin{tabular}{ c c c c | c c }
\hline\hline
Parameter & Value & Error & In 9b? & $\rm OC$ & $\Delta {\rm OC}$ \\
\hline
% The values in this table were re-checked by Colin 2020.05.
\Teff\,(K)     & $9495$ & $104$ & & & \\
\logg          & $4.17$ & $0.17$ & & & \\
\vsini & $114.9$ & $3.4$ & & & \\
%$\log_{10}$(Fe/H) & $-4.43$ & $0.14$ & & & \\
\vmic  & $2.02$    & $0.64$ & & & \\
\hline
H     & 0.00   &          & H\,I & & \\
He    & -0.89  & 0.20$^1$ &  & $-1.14$ & $[-2.00;-0.84]$\\
%%Li    & <-8.0  &     $^1$  & $$ & $[;]$\\
Li    & <-8.0$^1$ &       &  & $$ & \\
C     & -3.61  & 0.08$^4$ &  & $-3.32$ & $[-3.55; -3.08]$\\
N     & -3.88  & 0.20$^1$ &  & $$ & \\
O     & -3.19  & 0.13$^5$ & \newc{O\,I} & $-3.33$ & $[-3.41;-3.22]$\\
Na    & -5.60  & 0.20$^1$ & Na\,I  & $-5.79$ & $[-6.13;-5.51]$\\
Mg    & -4.34  & 0.16$^5$ & Mg\,I  & $-4.58$ & $[-5.81;3.80]$\\
Al    & -5.32  & 0.30$^2$ &  & $-5.65$ & $[6.19;-5.35]$\\
Si    & -4.33  & 0.16$^5$ &  & $-4.51$ & $[-6.36; -3.58]$\\
S     & -4.53  & 0.30$^2$ &  & $-4.39$ & $[-4.73;-4.08]$\\
Ca    & -5.45  & 0.13$^5$ & Ca\,II  & $-5.60$ & $[-6.23;-4.86]$\\
Sc    & -8.85  & 0.15$^4$ & Sc\,II  & $-8.93$ & $[-9.68;-7.48]$\\
Ti    & -6.93  & 0.05$^4$ & Ti\,II  & $-7.04$ & $[-7.64;-6.49]$\\
V     & -7.73  & 0.40$^1$ &  & $-7.78$ & $[-8.19;-7.22]$\\
Cr    & -6.27  & 0.13$^4$ & Cr\,II  & $-6.20$ & $[-6.81;-5.15]$\\
Mn    & -6.88  & 0.40$^2$ &  & $-6.49$ & $[-6.83;-5.18]$\\
Fe    & -4.43  & 0.14$^6$ & Fe\,I, Fe\,II  & $-4.56$ & $[-5.08; -3.81]$\\
Ni    & -5.74  & 0.14$^2$ &  & $-5.77$ & $[-6.11;-5.06]$\\
Sr    & -9.16  & 0.15$^1$ &  & $-9.28$ & $[-9.28;-9.28]$\\
Y     & -9.36  & 0.30$^2$ & Y\,II  & $-9.56$ & $[-9.93;-8.64]$\\
Zr    & -9.01  & 0.25$^1$ &  & $$ & \\
Ba    & -9.78  & 0.15$^2$ &  & $-9.38$ & $[-10.1;-8.17]$\\
%Nd    &        &          &  & $-9.69$ & $[-9.69;-9.69]$\\
%Pr    &        &          &  & $-7.57$ & $[-7.57;-7.57]$\\
%Zn    &        &          &  & $-7.37$ & $[-7.94;-6.79]$\\
\hline
\end{tabular}
\\\begin{flushleft}
\emph{Notes. }Detections of atomic species in the atmosphere of the planet KELT-9b can be found in \citet[][]{YanHenning2018} for H$\alpha$; \citet[][]{Cauleyetal2019} for H$\alpha$, H$\beta$, Mg\,I; \citet[][]{Hoeijmakersetal2019} for Na\,I, Mg\,I, Sc\,II, Ti\,II, Cr\,II, Fe\,I, Fe\,II; \citet[][]{Turneretal2020} for H$\alpha$ and Ca\,II; \newc{and \citet{Borsaetal2022} for O\,I}.
\end{flushleft}
\label{tab:param-abuns}
\end{table}

\section{Conclusions}

\newb{We have measured the currently most comprehensive photospheric composition for the star KELT-9 (HD\,195689), including for the first time all \newc{nine }elements previously detected in its rapidly evaporating ultra-hot Jupiter, KELT-9b.}

\newb{We observe no significant deviation in any elemental abundance with respect to a sample of similar stars in several open clusters. In spite of the relatively slow mixing processes in the envelopes of A0 stars such as KELT-9, this lack of anomalies is consistent with our calculations for the photospheric contamination fraction using observationally derived evaporation rates for KELT-9b, where we find $f_{\rm ph}=-5.98$ for hydrogen-dominated material and $f_{\rm ph}=-7.52$ for iron.}

\newb{The results also rule out any recent ($\lesssim 10^{4}\,$yr) ingestion by the star of rocky Earth-mass moons lost by the planet. This is in line with calculations \newc{of satellite orbital stability }in the literature, which suggest that if the hot Jupiter arrived at its present orbit shortly after formation, any such moons would likely have been lost to the star while the system was still a fraction of its current ${\sim}1\,$Gyr age.}

\newb{Our analysis can serve as a template for seeking chemical fingerprints of planetary mass transfer or ingestion in the photospheres of other early-type stars, which can be relatively easily polluted by recently accreted material.}

\section*{Acknowledgements}
\newc{The authors thank the anonymous referee and Luca Fossati for their constructive feedback on the manuscript. The authors contributed equally to this work. }
MK gratefully acknowledges funding from the European Union's Horizon 2020 research and innovation programme under the Marie Sklodowska-Curie grant agreement No 753799. ASJ thanks the UK Marshall Commission for financial support. 
The Flatiron Institute is funded by the Simons Foundation.
This research is funded by the Gordon and Betty Moore Foundation through Grant GBMF7392. This research was supported in part by the National Science Foundation under Grant No. NSF PHY-1748958. Based on observations obtained through the Gemini Remote Access to CFHT ESPaDOnS Spectrograph (GRACES) at the international Gemini Observatory, a program of NSF’s NOIRLab, which is managed by the Association of Universities for Research in Astronomy (AURA) under a cooperative agreement with the National Science Foundation on behalf of the Gemini Observatory partnership: the National Science Foundation (United States), National Research Council (Canada), Agencia Nacional de Investigaci\'{o}n y Desarrollo (Chile), Ministerio de Ciencia, Tecnolog\'{i}a e Innovaci\'{o}n (Argentina), Minist\'{e}rio da Ci\^{e}ncia, Tecnologia, Inova\c{c}\~{o}es e Comunica\c{c}\~{o}es (Brazil), and Korea Astronomy and Space Science Institute (Republic of Korea). \newb{This work was enabled by observations made from the Gemini North telescope, located within the Maunakea Science Reserve and adjacent to the summit of Maunakea. We are grateful for the privilege of observing the Universe from a place that is unique in both its astronomical quality and its cultural significance.}

\section*{Data Availability}

The Gemini/GRACES spectroscopic observations of KELT-9 (HD\,195689) are stored and openly accessible in the Gemini Observatory Archive\footnote{\texttt{https://archive.gemini.edu}}. The FITS files for program GN-2017-DD-1 can be found by searching under PI name ``Kama'' with the target ``HD\,195689''.  The CAMstars software instrument with its associated data is available on GitHub\footnote{\texttt{https://github.com/adamjermyn/CAMstars}.}.

\appendix

%%%%%%%%%%%%%%%%%%%%%%%%%%%%%%%%%%%%%%%%%%%%%%%%%%
\bibliographystyle{mnras}
\bibliography{kelt9}

% Don't change these lines
\bsp	% typesetting comment
\label{lastpage}
\end{document}